\documentstyle[aps,preprint,epsfig]{revtex}
\def\beas{\begin{eqnarray*}}
\def\eeas{\end{eqnarray*}}
\def\bq{\begin{equation}}
\def\eq{\end{equation}}
\def\ben{\begin{enumerate}}\def\een{\end{enumerate}}
\newcommand{\be}{\begin{equation}}
\newcommand{\ee}{\end{equation}}
\newcommand{\bear}{\begin{eqnarray}}
\newcommand{\eear}{\end{eqnarray}}

\def\Tr{{\rm Tr}\, }

\def\parallel{| \hskip-0.03cm |}

\def\be{\begin{eqnarray}}
\def\ee{\end{eqnarray}}

\def\ln{{\rm ln}}

\def\A0{A_0}

\def\roughly#1{\mathrel{\raise.3ex\hbox{$#1$\kern-.75em%
\lower1ex\hbox{$\sim$}}}}

\def\plb #1 {Phys.~Lett.~B~{\bf #1}\, }
\def\np #1 {Nucl.~Phys.~{\bf #1}\, }
\def\prd #1 {Phys.~Rev.~D~{\bf #1}\, }
\def\prc #1 {Phys.~Rev.~C~{\bf #1}\, }
\def\prl #1 {Phys.~Rev.~Lett.~{\bf #1}\, }

\begin{document}
\tightenlines

\preprint{\parbox[b]{1in}{
\hbox{\tt PNUTP-03/A02}
\hbox{\tt SNUTP-03/007}
}}

\draft

\title{Higher order corrections to Color superconducting gaps}

\author{Deog Ki Hong$^{1,a}$, Taekoon Lee$^{2,b}$, Dong-Pil Min$^{3,c}$,
D. Seo$^a$, and Chaejun Song$^{4,c}$}

\vspace{0.05in}

\address{
$^{a}$Department of Physics, Pusan National University,
Pusan 609-735, Korea
\protect\\
$^{b}$Department of Physics,
KAIST, Taejon 305-701, Korea
\protect\\
$^c$School of Physics and CTP,
Seoul National University, Seoul 151-747, Korea
\protect\\   
\vspace{0.05in}
{\footnotesize\tt $^{1}$dkhong@pnu.edu,
$^2$tlee@muon.kaist.ac.kr,
$^{3}$dpmin@phya.snu.ac.kr,
$^4$chaejun@phya.snu.ac.kr}
}

\vspace{0.1in}

\date{\today}

\maketitle

\begin{abstract}
We find a (nonlocal) gauge where the wavefunction renormalization
constant does not get any corrections for all momenta
in the hard-dense loop approximation.
In this gauge, we solve the Schwinger-Dyson equations for
the diquark condensate in dense QCD to calculate the Cooper pair
gap. We determine not only the exponent but also the prefactor of the
gap in a gauge independent way.
We find that the higher order corrections increase the gap
only by about 1.6 times to the leading order gap at Coulomb
gauge.
\end{abstract}

\pacs{PACS numbers: 12.38.Aw, 12.38.Mh, 11.15.Ex}

Matter at extreme density is known to be a color
superconductor~\cite{bailin}, quarks forming Cooper pairs to
open a gap at the Fermi momenta. The properties of color
superconductor are intensively studied recent years at
asymptotic density~\cite{review}, where quarks interact weakly
due to asymptotic freedom in quantum chromodynamics (QCD).
However, the study beyond the leading order
has been hindered by calculational difficulties.

Since superconductors are characterized by the Cooper-pair gap or
the minimum energy to excite a pair of a quark and a hole at the
Fermi momenta, it is quite important to determine the size of the
gap accurately. By solving the Schwinger-Dyson (SD) equations, the
gap at the asymptotic density has been found to be
\begin{eqnarray}
\Delta_0\simeq b{\mu\over g_s^5}
\exp\left(-{c\over g_s}\right)
\label{gap}
\end{eqnarray}
where the $1/g_s$ power in the
exponent is due to the long-range interaction in dense QCD.
The constant $c$ in the exponent is gauge-independent and is
determined by the long-range interaction mediated
by the magnetic gluons, unscreened at high baryon density.
In the hard-dense loop
(HDL) approximation, it is known to be
$c=3\pi^2/\sqrt{2}$~\cite{son}.
The short-range interaction mediated by the Debye-screened
electric gluons contributes only to the prefactor.
Among the negative fifth power of the coupling in the prefactor,
the third is due to the screened electric gluons, while the
magnetic gluons contribute the remaining.
In the leading order analysis of the SD equation~\cite{Hong:1999fh,Hong:1998tn}
the numerical prefactor $b$ is found to be
\begin{equation}
b=2^7\pi^4\left({2\over N_f}\right)^{5/2}e^{3\xi/2+1}~.
\label{prefactor}
\end{equation}
where $N_f$ is the number of quark flavors and $\xi$ is the gauge-fixing parameter.

The result Eq.~(\ref{prefactor}) is gauge-dependent,
since all subleading contributions
that lead to logarithmic divergences as in the BCS
superconductivity are not taken into account.
Being a physical observable, the gap $\Delta_0$ should be
gauge-independent. However, if one sums the contributions partially,
the result is often gauge-dependent.
We therefore need to calculate the subleading corrections to obtain the
gauege-independent gap.
Some of the subleading corrections like
the finiteness of the quasi-quark life in medium~\cite{Manuel:2000nh},
the running effect of the strong coupling~\cite{Beane:2000hu}, and the
quark self-energy~\cite{Brown:1999aq,Wang:2001aq}
have been studied.

In this paper we determine accurately the contributions to the prefactor,
coming from the vertex corrections and the wavefunction renormalization
for quarks, since these are the remaining gauge-dependent contributions
to the prefactor and, if summed up, we should get a
gauge-independent prefactor.

We first try to find a (nonlocal) gauge~\cite{Georgi:1989cd}
where the quark
wavefunction is not renormalized for all momenta, $Z(p)=1$.
In this gauge we then solve the SD equations to find the Cooper
pair gap, paying special attention to the Ward-Takahashi identity.
This way of calculating higher order corrections in the SD analysis has been
proved extremely useful in dynamical mass generation of (2+1)-dimensional
quantum electrodynamics and in others, since the higher order corrections vanish
exactly~\cite{Ebihara:1994wm,Nash:1989xx}.

Unlike ordinary electron superconductors, without additional interaction
the quark-quark scattering is attractive
in the color antitriplet channel,
where the color flux energy is lowered.
By the Cooper theorem, diquark operators of opposite momenta therefore
develop a condensate in quark matter, breaking the color gauge symmetry,
\begin{equation}
\left<\psi(x)\bar\psi_c(x)\right>
=K(p_F),
\end{equation}
where the charge-conjugated field is defined as
$\left(\psi_c\right)_{i}(x)=C_{ij}\bar\psi_{j}(x)$.
The matrix $C$
satisfies $C^{-1}\gamma_{\mu}C=-\gamma_{\mu}^T$ and
$i,j$ are Dirac indices.

To calculate the condensate, we introduce a Nambu-Gorkov field
$\Psi(x)\equiv(\psi(x),\psi_c(x))^T$.
The inverse propagator for the Nambu-Gorkov field is then given as
\begin{equation}
S^{-1}(p)= -i\pmatrix{a(p)\left[(p_0+\mu)\gamma^0+b(p)\not\!\vec p\,\right] &
       -\Delta(p) \cr
-\gamma^0\Delta^{\dagger}(p)\gamma^0&
a(p)\left[(p_0-\mu)\gamma^0+b(p)\not\!\vec p\,\right]\cr},
\label{inverse}
\end{equation}
where
$a(p), b(p)$ denote the wave function renormalization constants
and $\Delta(p)$  denotes the Cooper-pair gap at a momentum $p$.
Notice that the time and spatial components of quark wavefunction
renormalize differently since the Lorentz symmetry is broken in quark matter.

\begin{figure}
\epsfxsize=5in
\centerline{\epsffile{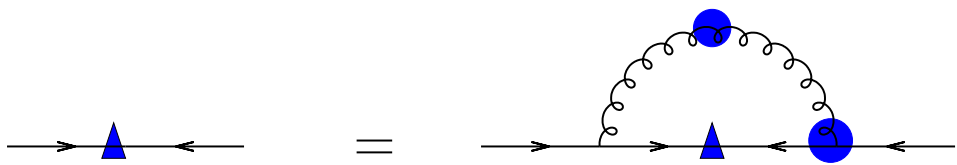}}
\caption{The triangle denotes the Cooper-pair gap and the blobs
higher-order corrections.}
 \label{fig1}
\end{figure}

We study the Schwinger-Dyson equation for
the Nambu-Gorgov propagator, following the notations in~\cite{Hong:1999fh}.
The gap equation in dense QCD takes a following form (See Fig.~\ref{fig1});
\begin{eqnarray}
\Delta(p_0)&=& {g_s^2\over c^2}\int {\rm d}q_0
{\Delta(q_0)\over \sqrt{q_0^2+\Delta^2}}
\left[\left(1+\eta\right)\ln\left({\mu\over |p_0-q_0|}
\right)+\ln\,b+\zeta\right].
\label{gapf}
\end{eqnarray}

The first term in Eq.~(\ref{gapf}) comes from the magnetic gluons and gives
the leading contribution. The constant term, $\ln\,b$, is due to the
Debye screened electric gluons and the gauge fixing terms.
The higher-order terms, denoted as $\eta$ and $\zeta$,  are in general
given in powers of coupling constant and energy as
$g_s^l\,
 \left[\frac{\Delta, q_0}{\mu}\right]^m
\left[\ln\left(\frac{\Delta, q_0}{\mu}\right)\right]^n.$
In our calculation $q_0\sim \Delta$.
The higher-order corrections are in general suppressed by powers of coupling
constant. However, since the gap depends on the coupling, the logarithmic
corrections might make the higher-order corrections comparable
to the leading term. We examine the two-loop corrections to the gap
equation in detail.

\begin{figure}
\epsfxsize=5in
\centerline{\epsffile{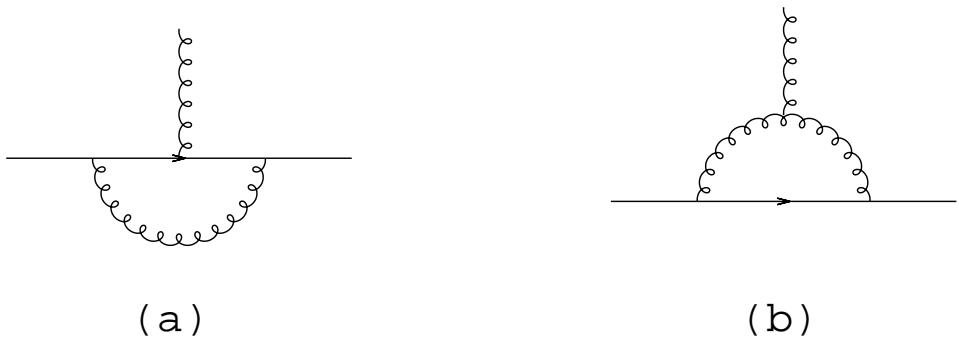}}
\caption{The solid line denotes quarks and the curly lines
gluons.}
 \label{fig2}
\end{figure}

In general, the SD equations are infinitely coupled equations for
the 1PI functions. However, some of them are related by
Ward-Takahashi (WT) identities. The identity we are going to use
in solving the gap equation is one that relates
the quark two-point functions with the vertex functions,
\begin{eqnarray}
& &\partial_z^{\mu}\left<j_{\mu}^a(z)\psi(x)\bar\psi(y)\right>\nonumber\\
=& &
\left<\partial^{\mu}j_{\mu}^a(z)\psi(x)\bar\psi(y)\right>
-\delta(z-x)\left<T^a\psi(x)\bar\psi(y)\right>+\delta(z-y)
\left<\psi(x)\bar\psi(y)T^a\right>.
\end{eqnarray}
The identity reads in the momentum space as
\begin{eqnarray}
& & T^a ~a(p)\left[(p_0+\mu)\gamma^0+b(p)\not\!\vec p\right]-
a(p^{\prime})\left[(p^{\prime}_0+\mu)\gamma^0+b(p^{\prime})\not\!
\vec p^{\prime}\right]T^a \nonumber\\
& & =(p-p^{\prime})_{\mu}\Lambda^{\mu}(p,p^{\prime})T^a\,
+\Gamma^a(p,p^{\prime};-p-p^{\prime}),
\end{eqnarray}
where
\begin{equation}
\Gamma^a(p,p^{\prime};k)\delta(k+p+p^{\prime})=
\int_{z,x,y}e^{i(z\cdot k+x\cdot p+y\cdot p^{\prime})}
\left<\partial^{\mu}j_{\mu}^a(z)\psi(x)\bar\psi(y)\right>.
\end{equation}
The one-loop vertex correction has two parts (See Fig.~{\ref{fig2}}).

As in quantum electrodynamics,
the first part (Fig.~2a) is related to the correction to the wavefunction
renormalization constant as
\begin{eqnarray}
(p-p^{\prime})_{\mu}\Lambda^{(a)\mu}(p,p^{\prime})
=a(p)\left[(p_0+\mu)\gamma^0+b(p)\not\!\vec p\right]-
a(p^{\prime})\left[(p^{\prime}_0+\mu)\gamma^0+b(p^{\prime})\not\!
\vec p^{\prime}\right]\nonumber,
\end{eqnarray}
where we suppressed the color indices.
We see that $\Lambda^{(a)\mu}=\gamma^{\mu}$ for the nonlocal gauge
where the wavefunction renormalization constants are $a(p)=1=b(p)$.
(This result is similar to the condition obtained by Kugo and Mitchard
in the SD analysis of chiral symmetry breaking in QCD, using vector
Ward identities~\cite{Kugo:1992pr}.)
Therefore, if we find a gauge where $a(p)=1=b(p)$,
not only the corrections from the wavefunction renormalization
but also those from the vertex corrections from Fig.~2a are absent. This
simplifies significantly the calculation in the higher order corrections
to the Cooper-pair gap, since
the vertex correction to the gap, coming
from the second diagram, Fig.~2b, does not contribute to the
prefactor, $b$, in Eq.~(\ref{gap}).
To see this, we note the leading contributions to the vertex comes
from when both the internal and the external gluon lines are magnetic.
Furthermore, the main contribution to the gap occurs when the external lines
carry small momenta.
For small external gluon and quark momenta
($q, |\vec p|-\mu,|\vec{p}^{\prime}|-\mu \sim\Delta$),
the vertex correction due to the diagram Fig.~2b is given as,
using the high density effective theory of QCD~\cite{Hong:1998tn},
\begin{eqnarray}
I_{\mu}^a\simeq\,g_s^3
\int_l{l\cdot V\over l_{\parallel}^2+\Delta^2}{
\gamma_0f^{abc}T^bT^c\left[\left(c_1l_{\parallel}^i+
c_2l_{\perp}^i+c_3p^i+c_4p^{\prime i}\right)
g_{\mu i}+c_5\left(2l-p-p^{\prime}\right)_{\mu}\right]\over
\left[\left|\vec l-\vec p\right|^2+\pi M^2\left|l_0-p_0\right|/
\left(2\left|\vec l-\vec p\right|\right)\right]\left[
\left|\vec l-\vec {p^{\prime}}\right|^2+\pi M^2\left|l_0-p^{\prime}_0\right|/
\left(2\left|\vec l-\vec {p^{\prime}}\right|\right)
\right]},
\end{eqnarray}
where $c_1$ and $c_5$ are $1+O\left(\Delta^2/l_{\perp}^2\right)$ while
all other $c_i$'s are $O\left(\Delta^2/l_{\perp}^2\right)$.
The screening mass $M$ is given as $g_s\mu\sqrt{N_f}/(2\pi)$ for
$N_f$ light quarks in the hard-dense-loop (HDL) approximation.
Therefore, we find upon integration the correction becomes
\begin{eqnarray}
I_{\mu}^a\sim i\,g_s^3\gamma_0V_{\mu}T^a\left({\Delta\over M}\right)^2\ln\left(
{\Delta/\mu}\right),
\end{eqnarray}
which is indeed negligible compared to the constant term,
$\ln\,b$, in the gap equation Eq.~(\ref{gapf}).

Now, we look for a nonlocal gauge where the wave function
constants remain unrenormalized, $a(p)=1=b(p)$.   At this gauge,
the vertex is bare, $\Lambda^{\mu}=\gamma^{\mu}$,
and the SD equation leads to an equation for the (nonlocal)
gauge-fixing parameter, given as
\be
\int d^4q D_{\mu\nu}(q-p)\Tr
(\gamma^0\gamma^\mu\Lambda^+_q\gamma^0\gamma^\nu\Lambda_p^-)
\frac{(q_0+|\vec{q}|-\mu)}{q_0^2-(|\vec{q}|-\mu)^2-\Delta^2}=0,
\label{eq1}
\ee
where the antiquarks are projected out by
the positive and negative energy state projectors, defined as
\begin{equation}
\Lambda^{\pm}_p\equiv {1\over2}\left(1\pm{\vec \alpha\cdot\vec p\over
\left|\vec p\right|}\right).
\end{equation}

The (Higgsed) gluon propagator is given in the HDL approximation as
\footnote{For two flavor color superconductors,
some of the gluons are not Higgsed. However, since the Meissner effect is
subleading, its inclusion does not affect our calculation.}
\be
D_{\mu\nu}(k)=A\, O_{\mu\nu}^{(1)} +B\, O_{\mu\nu}^{(2)}+
C\, O_{\mu\nu}^{(3)}\,,
\ee
where in the weak coupling limit, $|k_0|\ll|\vec k|$,
\begin{eqnarray}
A={\left|\vec k\right|\over \left|\vec k\right|^3+M_0^2\Delta+\pi M^2
\left|k_4\right|/2}\,,\quad
B={1\over k_4^2+{\vec k}^2+2M^2}\,,\quad
C={\xi \over k_4^2+{\vec k}^2},
\end{eqnarray}
where $M_0$ is Higgs-like gluon mass
$\sim g_s\mu /(2\pi)$~\cite{Hong:1999fh}.
The polarization tensors are defined as~\cite{Hong:1999fh}
\begin{eqnarray}
O^{(1)}=P^{\perp}+{(u\cdot k)^2\over (u\cdot k)^2-k^2}P^u,\nonumber\\
O^{(2)}=P^{\perp}-O^{(1)},\quad O^{(3)}=P^{\parallel},
\end{eqnarray}
where $u_{\mu}=(1,\vec 0)$ and
\begin{eqnarray}
P^{\perp}_{\mu\nu}=g_{\mu\nu}-{k_{\mu}k_{\nu}\over k^2},\quad
P^{\parallel}_{\mu\nu}={k_{\mu}k_{\nu}\over k^2},\nonumber\\
P^u_{\mu\nu}={k_{\mu}k_{\nu}\over k^2}-
{k_{\mu}u_{\nu}+u_{\mu}k_{\nu}\over (u\cdot k)}
+{u_{\mu}u_{\nu}\over (u\cdot k)^2} k^2.
\end{eqnarray}
Contracting the tensors, we get
\bear
O_{\mu\nu}^1 \Tr
(\gamma^0\gamma^\mu\Lambda^+_q\gamma^0\gamma^\nu\Lambda_p^-)&=&
-2 (1+t) \frac{q^2+p^2 -qp(1+t)}{q^2+p^2-2qpt} \nonumber\\
O_{\mu\nu}^2 \Tr
(\gamma^0\gamma^\mu\Lambda^+_q\gamma^0\gamma^\nu\Lambda_p^-)&=&
1-t \nonumber\\
O_{\mu\nu}^3 \Tr
(\gamma^0\gamma^\mu\Lambda^+_q\gamma^0\gamma^\nu\Lambda_p^-)&=&
-(1-t)\frac{(p+q)^2}{q^2+p^2-2pqt}
\eear
where $p=|\vec{p}|$, $q=|\vec{q}|$, and
$t=\vec p\cdot \vec q/(pq)$.

Now, we assume the (nonlocal) gauge fixing parameter $\xi(k)$
has only temporal dependence, that
is, $\xi(k)\simeq\xi(k_0)$, and would like to have (\ref{eq1})
satisfied already at angular integration.
This gives
\be
D+E+F=0
\label{eq2}
\ee
where
\bear
D&=& q^2 \int_{-1}^1 dt\, A(p_4-q_4,\vec{p}-\vec{q})\left[ -2 (1+t) \frac{q^2+p^2
-q p(1+t)}{q^2+p^2-2qpt}\right]\nonumber\\
&\approx& \frac{2}{3}\,\ln \left[\frac{(2\mu)^3}{
(q-p)^3 + M_0^2\Delta +\pi M^2|p_4-q_4|/4} \right]\nonumber \\
E&=&q^2 \int_{-1}^1 dt\, B(p_4-q_4,\vec{p}-\vec{q}) (1-t)\approx 1 \nonumber \\
F&=&q^2 \int_{-1}^1 dt\,
C(p_4-q_4,\vec{p}-\vec{q})
\left[-(1-t)\frac{(p+q)^2}{q^2+p^2-2pqt}\right]\nonumber \\
&\approx&-\xi\, \ln\left[ \frac{ (2\mu)^2}{(p-q)^2 +|p_4-q_4|^2}\right].
\eear
In this calculation, we take  $p,q\simeq\mu$,
since the quarks are near the Fermi surface.
We find the solution of  Eq (\ref{eq2})
\be
\xi\approx \frac{\frac{2}{3} \ln \frac{(2\mu)^3}{ M_0^2\Delta +
\pi M^2|p_4-q_4|/2}}{\ln \frac{ (2\mu)^2}{|p_4-q_4|^2}}
\approx \frac{1}{3},
\label{gauge}
\ee
with $|p_4-q_4|\sim \Delta$.

Plugging Eq.~(\ref{gauge}) into the leading-order
expression for the gap Eq.~(\ref{prefactor}), we get
\begin{equation}
\Delta_0=2^7\pi^4\left({2\over N_f}\right)^{5/2}e^{3/2}
\exp\left(-{3\pi^2\over\sqrt{2}g_s}\right).
\end{equation}
We see that the vertex and the wave-function correction increase
the gap by about two thirds, compared
to the leading-order gap at the Coulomb gauge~\cite{sw,pr},
\begin{equation}
\Delta_0=e^{0.5}\Delta_0(\xi=0).
\end{equation}

In conclusion, we find that the wave-function renormalization vanish
at a nonlocal gauge-fixing parameter $\xi\simeq 1/3$. By the
Ward-Takahashi identity, the QED-like vertex correction vanishes
at this gauge as well.
Since the remaining one-loop vertex corrections
are suppressed for small external momenta, which has a major
contribution to the gap equation,
both the wave-function correction and the vertex correction to
the prefactor of the gap
vanishes at this nonlocal gauge. As the gap is independent of the choice
of gauge, we conclude that the next-to-leading order corrections due to
the vertex renormalization and the wavefunction renormalization
increase the leading-order gap at the Coulomb gauge by about two thirds.

\eject

\acknowledgments
One of us (DKH) wishes to thank D. Rischke for discussions.
The work of DKH was supported in part by
the academic research fund of Ministry of Education,
Republic of Korea, Project No. KRF-2000-015-DP0069.
The work of DKH and DPM was supported in part
by KOSEF grant number R01-1999-000-00017-0. The work of
DPM and CS was supported in part by KRF 2001-015-DP0085 and BK21
projects of Ministry of Education, Republic of Korea.
TL was supported in part by BK21 Core Project.

\eject

\vskip 0.2in

\end{document}